\documentclass[12pt]{article}

\input{epsf.tex}

\newcommand{\E}{{\cal{E}}}

\newcommand{\I}{{\rm i}}
\renewcommand{\a}{\alpha}
\renewcommand{\d}{{\rm d}}

\newcommand{\be}{\begin{equation}}
\newcommand{\ee}{\end{equation}}
\newcommand{\bea}{\begin{eqnarray}}
\newcommand{\eea}{\end{eqnarray}}

\newcommand{\nabv}{\mbox{\boldmath$\nabla$}}

\def\J#1#2#3#4{#1 {\it #2} {\bf #3} #4}

\def\PRD{{\it Phys. Rev.} D}

\def\JMP{\it J. Math. Phys.}
\def\CQG{\it Class. Quantum Grav.}

\begin{document}

\title{Singular sources in the Demia\'nski--Newman spacetimes}
\author{V S Manko$^1$, J Mart\'\i n$^2$ and E Ruiz$^2$}
\date{}
\maketitle

\vspace{-1cm}

\begin{center}
$^1$ Departamento de F\'\i sica,\\ Centro de Investigaci\'on y de
Estudios Avanzados del IPN,\\ A.P. 14-740, 07000 M\'exico D.F.,
Mexico\\ $^2$ Departamento de F\'{i}sica Fundamental, Universidad
de Salamanca,\\ 37008 Salamanca, Spain
\medskip
\end{center}

\begin{abstract}
The analysis of singular regions in the NUT solutions carried out
in the recent paper (Manko and Ruiz, 2005 {\it Class. Quantum
Grav.} {\bf 22}, 3555) is now extended to the Demia\'nski--Newman
vacuum and electrovacuum spacetimes. We show that the effect which
produces the NUT parameter in a more general situation remains
essentially the same as in the purely NUT solutions: it introduces
the semi--infinite singularities of infinite angular momenta and
positive or negative masses depending on the interrelations
between the parameters; the presence of the electromagnetic field
additionally endows the singularities with electric and magnetic
charges. The exact formulae describing the mass, charges and
angular momentum distributions in the Demia\'nski--Newman
solutions are obtained and concise general expressions
$P_n=(m+\I\nu)(\I a)^n$, $Q_n=(q+\I b)(\I a)^n$ for the entire set
of the respective Beig--Simon multipole moments are derived. These
moments correspond to a unique choice of the integration constant
in the expression of the metric function $\omega$ which is
different from the original choice made by Demia\'nski and Newman.
\end{abstract}



\newpage

\section{Introduction}

The Demmia\'nski--Newman (DN) stationary axisymmetric spacetimes
\cite{DNe} are (i) the Kerr solution \cite{Ker} endowed with the
NUT parameter \cite{NTU} (the pure vacuum case), and (ii) the
Kerr--Newman solution \cite{NCC} endowed with the NUT parameter
and magnetic charge (the case of electrovacuum). The canonical
Papapetrou form of the general DN class was obtained in \cite{AGM}
as the simplest $N=1$ specialization of the extended
multi--soliton electrovacuum metric \cite{RMM}.

The presence of the non--vanishing NUT parameter makes the DN
spacetimes asymptotically non--flat. This complicates the analysis
of the respective mass and angular momentum distributions since,
as was shown in \cite{MRu}, the singular semi-infinite regions
developed by the NUT parameter have in general infinite angular
momenta and non--zero masses. Moreover, the presence of the
electromagnetic field requires additional consideration of the
distributions of electric and magnetic charges in these singular
regions. As an interesting aspect of the analysis of DN solutions
one may consider establishing the correspondence between the
multipole moments of the asymptotically NUT spacetimes calculated
on a 3--manifold of trajectories of the timelike Killing vector
with the aid of Simon's procedure \cite{Sim} and the physical
characteristics of the entire four--dimensional metric.

The plan of our paper is the following. In section~2 we consider
the Ernst complex potentials and corresponding metric functions
defining the DN spacetimes; here in particular we obtain concise
formulae for the entire set of the Beig--Simon (BS) multupole
moments associated with these solutions. Section~3 is devoted to
the analysis of the mass and angular momentum distributions in the
DN vacuum spacetimes, also known in the literature under the name
Kerr--NUT metric. The more general case involving the NUT
parameter and the electromagnetic field is analyzed in section~4.
Lastly, concluding remarks are given in section~5.

\section{The Ernst potentials, metric functions and BS
multipole moments of the DN solutions}

It is commonly known that the stationary axisymmetric
electrovacuum solutions of the Einstein--Maxwell equations are
most easily presentable and analyzed if one uses, on the one hand,
the Papapetrou canonical line element
\begin{equation}\label{papa}
\d s^2= f^{-1}[e^{2\gamma} (\d\rho^2 + \d z^2) + \rho^2
\d\varphi^2]- f( \d t - \omega \d\varphi)^2,
\end{equation}
where the coefficients $f$, $\gamma$, $\omega$ are functions of
the cylindrical coordinates $\rho$, $z$ only, and, on the other
hand, the Ernst formalism developed in the papers \cite{Ern1,Ern2}
according to which the knowledge of two complex potentials $\E$
and $\Phi$ satisfying the Ernst equations \cite{Ern2}
\bea\label{EEq} ({\rm Re}\E+\Phi\bar\Phi)\Delta\E=
(\nabv\E+2\bar\Phi\nabv\Phi)\nabv\E, \nonumber\\ ({\rm
Re}\E+\Phi\bar\Phi)\Delta\Phi=
(\nabv\E+2\bar\Phi\nabv\Phi)\nabv\Phi \eea (a bar over a symbol
denotes complex conjugation, $\nabv$ and $\Delta$ are the
3--dimensional gradient and Laplace operators, respectively) is
sufficient for the reconstruction of the metric coefficients
entering (\ref{papa}), together with the electric and magnetic
components of the electromagnetic 4--potential.

Demia\'nski and Newman originally derived their solutions by
employing some specific complex coordinate transformations not
preserving Papapetrou's form of the line element; the Ernst
potentials defining the general class of DN metrics were obtained
later in \cite{AGM} using the extended multi--soliton
electrovacuum solution \cite{RMM} constructed with the aid of
Sibgatullin's method \cite{Sib,MSi}; these potentials have the
form \bea\label{DN_Ep} \E=\frac{\kappa x-m-\I(ay+\nu)}{\kappa
x+m-\I(ay-\nu)},
\quad \Phi=\frac{q+\I b}{\kappa x+m-\I(ay-\nu)}, \nonumber\\
x=\frac{1}{2\kappa}(r_++r_-), \quad
y=\frac{1}{2\kappa}(r_+-r_-), \nonumber\\
r_\pm=\sqrt{\rho^2+(z\pm\kappa)^2}, \quad
\kappa=\sqrt{m^2+\nu^2-a^2-q^2-b^2}, \eea where the arbitrary real
parameters $m$, $\nu$, $a$, $q$, $b$ can be associated,
respectively, with the mass, gravomagnetic monopole (NUT
parameter), angular momentum per unit mass, electric and magnetic
charges. In the absence of the electromagnetic field ($q=b=0$
$\Leftrightarrow \Phi=0$) one obtains from (\ref{DN_Ep}) the
expression for $\E$ defining the vacuum DN spacetimes. When the
NUT parameter $\nu$ is equal to zero, one arrives at the
black--hole solutions considered by Carter \cite{Car}.

The potentials (\ref{DN_Ep}) can be used for the calculation of
the corresponding BS multipole moments \cite{BSi,SBe,Sim}. For
this purpose it is advantageous to employ the
Hoenselaers--Perj\'es procedure \cite{HPe}, recently rectified by
Sotiriou and Apostolatos \cite{SAp}, which involves the axis
expressions of the Ernst potentials. For $\rho=0$, $z>{\rm
Re}(\kappa)$ we have \bea\label{DN_axis} e(z)\equiv\E(\rho=0,z)=
\frac{z-m-\I(a+\nu)}{z+m-\I(a-\nu)}, \nonumber\\
f(z)\equiv\Phi(\rho=0,z)= \frac{q+\I b}{z+m-\I(a-\nu)}. \eea Then,
passing to the potentials $\xi$ and $\eta$ via the formulae \be
e(z)=\frac{1-\xi}{1+\xi}, \quad f(z)=\frac{\eta}{1+\xi},
\label{xi} \ee and considering the expressions of the functions
$\tilde\xi=z\xi$ and $\tilde\eta=z\eta$ in the limit $z\to\infty$,
we obtain \bea\label{exp} \tilde\xi=\frac{(m+\I\nu)z}{z-\I a}=
(m+\I\nu)\sum\limits_{n=0}^{\infty}\frac{(\I a)^n}{z^n},
\nonumber\\ \tilde\eta=\frac{(q+\I b)z}{z-\I a}= (q+\I
b)\sum\limits_{n=0}^{\infty}\frac{(\I a)^n}{z^n}. \eea

Remarkably, it turns out that the coefficients in expansions
(\ref{exp}) coincide {\it exactly} with the actual BS multipole
moments because, as one trivially verifies, all quantities
$M_{ij}$, $Q_{ij}$, $S_{ij}$, $H_{ij}$ defined by formulae (23) of
\cite{SAp} (they describe the deviations of the coefficients in
(\ref{exp}) from the multipole moments) are equal to zero
identically. Hence we arrive at the following elegant explicit
expressions for the complex multipole moments: \be\label{mult}
P_n=M_n+\I J_n=(m+\I\nu)(\I a)^n, \quad Q_n=E_n+\I B_n=(q+\I b)(\I
a)^n, \ee where the real quantities $M_n$, $J_n$, $E_n$, $B_n$
are, respectively, the mass, angular momentum, electric and
magnetic multipole moments of the DN solution (\ref{DN_Ep}).
Formulae (\ref{mult}) generalize in a very natural and
straightforward way the Sotiriou--Apostolatos result \cite{SAp}
derived for the Kerr--Newman spacetime. The explicit expressions
for $M_n$, $J_n$, $E_n$ and $B_n$ are readily obtainable from
(\ref{mult}) by considering separately the even and odd moments:
\bea\label{MJ} M_{2k}=(-1)^{k}ma^{2k}, \quad
M_{2k+1}=(-1)^{k+1}\nu a^{2k+1}, \nonumber\\ J_{2k}=(-1)^{k}\nu
a^{2k}, \quad J_{2k+1}=(-1)^{k}m a^{2k+1}, \nonumber\\
E_{2k}=(-1)^{k}qa^{2k}, \quad E_{2k+1}=(-1)^{k+1}b a^{2k+1},
\nonumber\\ B_{2k}=(-1)^{k}b a^{2k}, \quad B_{2k+1}=(-1)^{k}q
a^{2k+1}, \quad k=0,1,\ldots \eea

We find it appropriate to make now several comments on the
multipole moments obtained. First of all, one should remember that
the BS moments, like the Geroch--Hansen multipole moments
\cite{Ger,Han}, are only well-defined for the asymptotically flat
spacetimes. In the stationary axisymmetric case these are
spacetimes whose metric coefficients in (\ref{papa}) have the
following behavior at spatial infinity: \be\label{asympt} f\to 1,
\quad \gamma\to 0, \quad \omega\to 0. \ee

Although the last condition on $\omega$ does not enter explicitly
into the procedure for the calculation of multipole moments
because the latter are defined on a specific 3--dimensional
manifold, it is normally taken into account by demanding that the
NUT parameter of the solution be equal to zero. As was already
remarked by Simon \cite{Sim}, the presence of the NUT parameter
and magnetic charge may cause the three- and four--dimensional
descriptions of a stationary electrovac solution topologically
incompatible.

Secondly, the multipole expansions define the corresponding Ernst
potentials uniquely, which is a consequence of the mathematical
theorems on analytic functions. On the other hand, the
corresponding metric functions $\gamma$ and $\omega$ are
constructed from the potentials $\E$ and $\Phi$ up to two
arbitrary real additive constants. The constant in $\gamma$ can be
easily adjusted to have the desired behavior of $\gamma$ even in
the presence of the NUT parameter; besides, $\gamma$ is involved
explicitly in the 3--dimensional manifold which is used for the
calculation of the multipole moments and hence can be always
defined uniquely by imposing  the asymptotic flatness condition.
The unique choice of the integration constant in the expression of
$\omega$ leading to $\omega\to 0$ at spatial infinity can be
realized only in the absence of the NUT parameter. When the
parameter NUT is present in the solution, the formal multipole
expansions of the corresponding Ernst potentials are still
possible, but the respective 4--dimensional spacetime is not
globally asymptotically flat for any choice of the integration
constant in $\omega$. Therefore, an important non--trivial
question arises of whether the multipole moments of the
asymptotically NUT spacetimes obtainable from the Ernst potentials
describe correctly the sources, and which is the precise choice of
the integration constant in $\omega$ corresponding to those
moments?

Thirdly, because it is clear that the analysis of sources in the
asymptotically NUT solutions cannot be restricted to only the
calculation of the multipole moments, one also needs a procedure
of the evaluation of physical quantities which would be applicable
to asymptotically non--flat spacetimes. The Komar integrals
\cite{Kom} then seem to be the best option as they had already
proved to be very efficient for treating the black--hole
properties in the external gravitational fields \cite{Tom1,BGM}.
In our recent paper \cite{MRu} we have applied the Komar integrals
to the analysis of the mass and angular momentum distributions in
the NUT spacetimes; the main result of that paper consists in
establishing the unique value of the integration constant entering
the expression of $\omega$ at which the total angular momentum is
a finite quantity and hence is in accordance with the multipole
structure. One may expect that an analogous criterion of the
choice of the integration constants can be worked out on the basis
of Komar integrals for the DN spacetimes too.

Let us proceed with the description of the DN spacetimes in the
Ernst picture by writing out the corresponding metric functions of
the electric $A_4$ and magnetic $A_3$ components of the
electromagnetic 4--potential worked out in \cite{AGM}:
\bea\label{mf_DN} f=\frac{\kappa^2(x^2-1)-a^2(1-y^2)}{(\kappa
x+m)^2+(ay-\nu)^2}, \quad {\rm
e}^{2\gamma}=\frac{\kappa^2(x^2-1)-a^2(1-y^2)}{\kappa
^2(x^2-y^2)}, \nonumber\\ \omega=2\nu(y+C_1)-
\frac{a(1-y^2)[2(m\kappa x-\nu ay+m^2+\nu^2)-q^2-b^2]}
{\kappa^2(x^2-1)-a^2(1-y^2)},\nonumber\\ A_4=\frac{q(\kappa
x+m)+b(\nu-ay)}{(\kappa x+m)^2+(ay-\nu)^2}, \nonumber\\
A_3=b(C_2-y)+\frac{(1-y)(ay+a-2\nu)[q(\kappa x+m)
+b(\nu-ay)]}{(\kappa x+m)^2+(ay-\nu)^2}. \eea Here $C_1$ and $C_2$
are two arbitrary real  (integration) constants whose concrete
values corresponding to the multipole moments (\ref{mult}) cannot
yet be pointed out. Formulae (\ref{mf_DN}) can be formally called
the entire family of DN spacetimes, and it is worthwhile
mentioning that the particular choice of $C_1$ and $C_2$ made in
\cite{AGM} does not lead to the geometry defined by the multipoles
(\ref{mult}).

It is convenient and instructive to consider the pure vacuum DN
spacetimes separately from the electrovacuum ones (as was done in
the original Demia\'nski--Newman paper \cite{DNe}): the simpler,
vacuum case provides a solid basis for understanding the general
structure of the DN sources, and we are passing to its
consideration in the next section.

\section{The mass and angular momentum distributions in the
DN vacuum subclass}

The vacuum subclass of DN spacetimes is normally known in the
literature as the ``genuine'' Demia\'nski--Newman metric. It is
also known under the name ``(combined) Kerr--NUT metric'' given to
it by Demia\'nski and Newman themselves because it contains one
additional (NUT) parameter compared to the Kerr solution
\cite{Ker}. Using formulae of the previous section let us write
out the Ernst potential $\E$ (the other potential $\Phi$ is equal
to zero identically) and corresponding metric functions
determining these particular vacuum spacetimes: \bea\label{DN_vac}
\E=f+\I\Omega=\frac{\kappa x-m-\I(ay+\nu)}{\kappa x+m-\I(ay-\nu)},
\quad \Omega=-\frac{2(\nu\kappa x+may)}{(\kappa x+m)^2+(\kappa
y-\nu)^2}, \nonumber\\ f=\frac{\kappa^2(x^2-1)-a^2(1-y^2)}{(\kappa
x+m)^2+(ay-\nu)^2}, \quad {\rm
e}^{2\gamma}=\frac{\kappa^2(x^2-1)-a^2(1-y^2)}{\kappa
^2(x^2-y^2)}, \nonumber\\ \omega=2\nu(y+C)-
\frac{2a(1-y^2)(m\kappa x-\nu ay+m^2+\nu^2)}
{\kappa^2(x^2-1)-a^2(1-y^2)},\nonumber\\
x=\frac{1}{2\kappa}(r_++r_-), \quad y=\frac{1}{2\kappa}(r_+-r_-),
\quad r_\pm=\sqrt{\rho^2+(z\pm\kappa)^2}, \nonumber\\
\kappa=\sqrt{m^2+\nu^2-a^2}, \eea where the real constant $C_1$
from (\ref{mf_DN}) is now called $C$, while the expression for
$\kappa$ does not contain the parameters $q$ and $b$ (cf. formula
(\ref{DN_Ep}) of the previous section).

When $C\ne\pm1$, $\nu\ne0$, the geometries (\ref{DN_vac}) are
characterized by two semi--infinite singular sources located on
the symmetry $z$--axis: $\rho=0$, $z>{\rm Re(\kappa)}$ (the upper
singularity), and $\rho=0$, $z<-{\rm Re(\kappa)}$ (the lower
singularity). In the particular cases $C=\pm1$, $\nu\ne0$, only
one semi--infinite singularity is present, which is analogous to
the pure NUT case \cite{MRu}. If $\kappa$ is a real non--zero
quantity, i.e. $m^2+\nu^2>a^2$, the symmetry axis is divided into
three parts: the upper region $I$ ($\kappa<z<\infty$), the
intermediate region $II$ ($|z|<\kappa$), and the lower region
$III$ ($-\infty<z<-\kappa$), see figure~1(a). For $a^2>m^2+\nu^2$,
$\kappa$ becomes a pure imaginary quantity
($\kappa=\I\sqrt{a^2-m^2-\nu^2}$), and only two different regions,
$I$ and $III$, will then be present on the $z$--axis
(figure~1(b)), the cut between the points $-\kappa$ and $+\kappa$
representing a superextreme central object.

Since our major interest lies in the semi--infinite sources which
are originated by the NUT parameter, in what follows we shall
mainly concentrate our consideration on the subextreme case
$m^2+\nu^2>a^2$ and we will pay less attention to the superextreme
case which can be considered as a sort of degeneration of the
sources structure (the region $II$ of the symmetry axis
disappears).

In the subextreme case the calculation of Komar quantities can be
suitably carried out with the aid of Tomimatsu's formulae obtained
in \cite{Tom2}: \bea M&=&-\frac{1}{4}\omega[\Omega(z=z_2)-
\Omega(z=z_1)], \nonumber\\ J&=&-\frac{1}{4}\omega(z_2-z_1)-
\frac{1}{8}\omega^2[\Omega(z=z_2)- \Omega(z=z_1)], \label{Tom_vac}
\eea where $M$ and $J$ define, respectively, the mass and angular
momentum of the part of the symmetry axis $[z_1,z_2]$ on which
$\omega$ takes the constant value.

The functions involved in formulae (\ref{Tom_vac}) are $\omega$
and $\Omega$ taken on the symmetry axis, and they have the
following form there:

{\it Region I}. This part of the $z$--axis is determined by
$\rho=0$, $z>\kappa$, or equivalently by $x=z/\kappa$, $y=1$.
Then, substituting these values of $x$ and $y$ into \ref{DN_vac},
we obtain
\be
\omega=2\nu(1+C), \quad \Omega=-\frac{2(\nu
z+ma)}{(z+m)^2+(a-\nu)^2}. \label{om_axis1} \ee

{\it Region II}. The intermediate region which is associated with
the central rotating body is defined by $\rho=0$, $|z|<\kappa$, or
$x=1$, $y=z/\kappa$, thus providing us with the following values
for $\omega$ and $\Omega$:
\be
\omega=\frac{2}{a}[m(\kappa+m)+\nu(aC+\nu)], \quad
\Omega=-\frac{2\kappa(maz+\kappa^2\nu)}
{(az-\kappa\nu)^2+\kappa^2(\kappa+m)^2}. \label{om_axis2} \ee

{\it Region III}. Here $\rho=0$, $z<-\kappa$, i.e. $x=-z/\kappa$,
$y=-1$, and we have
\be
\omega=2\nu(C-1), \quad \Omega=\frac{2(\nu
z+ma)}{(z-m)^2+(a+\nu)^2}. \label{om_axis3} \ee

We first apply formulae (\ref{Tom_vac}) to
(\ref{om_axis1})--(\ref{om_axis3}) for getting the masses $M_1$,
$M_2$, $M_3$ of the corresponding regions I, II, III, the result
is \bea &&M_1=-\frac{\nu(1+C)(a+\nu)}{2(\kappa+m)}, \quad
M_2=m+\frac{\nu(aC+\nu)}{\kappa+m}, \nonumber\\
&&M_3=\frac{\nu(1-C)(a-\nu)}{2(\kappa+m)}. \label{masses_vac} \eea
It is easy to verify that the total mass $M=M_1+M_2+M_3$ is equal
to $m$.

From (\ref{Tom_vac}) it follows that the entire angular momenta of
the semi--infinite singular regions are infinitely large
quantities since in the case of the region~I one has to put
$z_2=+\infty$, and in the case of the region~III $z_1=-\infty$.
Therefore, as in the paper \cite{MRu}, we shall calculate the
angular momentum $J_1(z_0)$ of the part $\kappa<z\le z_0$ of the
upper singular region, the total angular momentum $J_2$ of the
central body ($|z|\le\kappa$), and the part $-z_0\le z<-\kappa$ of
the lower semi--infinite singularity. Formulae (\ref{Tom_vac})
then give \bea &&\hspace{-1.5cm}J_1(z_0)=-\frac{\nu(1+C)}{2}\left[
z_0-\kappa+ \frac{\nu(1+C)(a+\nu)}{\kappa+m}\right]
+\frac{\nu^2(1+C)^2(\nu z_0+ma)} {(z_0+m)^2+(a-\nu)^2},
\nonumber\\
&&J_2=(a+C\nu)\left[m+\frac{\nu(aC+\nu)}{\kappa+m}\right],
\nonumber\\ &&\hspace{-1.5cm}J_3(z_0)=\frac{\nu(1-C)}{2}\left[
z_0-\kappa- \frac{\nu(1-C)(a-\nu)}{\kappa+m}\right]
-\frac{\nu^2(1-C)^2(\nu z_0-ma)} {(z_0+m)^2+(a+\nu)^2}.
\label{mom_vac} \eea

The total angular momentum of the part $|z|\le z_0$ of the
$z$--axis thus takes the form\footnote{There is a misprint in the
formula (15) of \cite{MRu}: the first expression in parentheses on
the right--hand side does not contain $\a$ and should be read as
$(z_0-2m)$.} \bea J(z_0&=&-C\nu(z_0-2m)+ma \nonumber\\
&&+\frac{\nu^2(1+C)^2(\nu z_0+ma)} {(z_0+m)^2+(a-\nu)^2}
-\frac{\nu^2(1-C)^2(\nu z_0-ma)} {(z_0+m)^2+(a+\nu)^2},
\label{mom_tot_vac} \eea and one can see that the only choice of
the constant $C$ leading to the finite value of $J$ in the limit
$z_0\to\infty$ is $C=0$. In particular, in the original DN
solution characterized by $C=-1$ the only (lower) semi--infinite
singularity carries an infinite angular momentum.

In the general case, the aggregate mass of two semi--infinite
singular sources is equal to
\be
M_{agg}=M_1+M_3=-\frac{\nu(aC+\nu)}{\kappa+m},
\label{mass_sing_vac} \ee and it can assume either positive or
negative values. When $C=0$, $M_{agg}$ is a positive quantity for
all non--zero $m$, $\nu$ and $a$. However, even in the cases when
$M_{agg}$ is positively defined, the negative mass is present in
the singular semi--infinite sources. Let us illustrate this by
taking as an example the original DN vacuum solution ($C=-1$). In
this case $M_{agg}=M_3=\nu(a-\nu)/(\kappa+m)$, so that $M_{agg}$
is positive for instance when $\nu>0$, $\nu<a<m$. In figure~2 we
have plotted the mass $M_3(z_0)$ of the part $-z_0\le z<-\kappa$
of the semi--infinite singularity, namely
\be
M_3(z_0)=\frac{\nu(a-\nu)}{\kappa+m} +\frac{2\nu(\nu
z_0-ma)}{(z_0+m)^2+(a+\nu)^2}, \label{m3_z0_sub} \ee as the
function of $z_0$, for the particular choice of the parameters
$m=5$, $\nu=3$, $a=4$. It follows that $M_3(z_0)$ is a
monotonously increasing function on the interval
$4.243<z_0<20.272$, taking its maximal value 0.681 at
$z_0=20.272$, and it is a monotonously decreasing function on the
interval $20.272<z_0<\infty$, reaching asymptotically the value
0.325, which means that the latter interval is entirely composed
of the negative mass.

Moreover, even in the special case $\nu=a$ when the total mass of
the DN semi--infinite singularity is equal to zero, the
qualitative picture of the mass distribution in the singularity is
similar to that shown in figure~2: the interval with the positive
mass is followed by the distribution of the negative mass. This is
illustrated in figure~3 for the particular choice of the
parameters $m=5$, $\nu=3$, $a=3$, the total mass of the
singularity reaching asymptotically the value $M_3=0$.

When the central body is a superextreme object, i.e., in the case
$a^2>m^2+\nu^2$, the structure of singularities preserves the main
characteristic features of the subextreme case: the aggregate
angular momentum of the singularities assumes a finite value only
when the constant $C$ is equal to zero (for the non--vanishing
$m$, $\nu$ and $a$), and the semi--infinite singular regions
involve negative masses. In figure~4 we have plotted the mass
distribution in the singularity of the original DN vacuum solution
($C=-1$) for two particular parameter sets which have common
values $m=2$, $\nu=1$ and differ in the value of $a$: in the first
set $a=4$ (figure~4(a)), and in the second $a=-4$ (figure~4(b)).

Since only the case $C=0$ is consistent with the multipole moments
defined by the Ernst potential (\ref{DN_vac}), below we shall give
the general expressions for the masses and angular momenta
exclusively for this superextreme case, the subindex~2 referring
to the central object: \bea &&M_1=-\frac{ma\nu}{m^2+(a-\nu)^2},
\quad M_2=\frac{m(m^2+\nu^2+a^2)^2}{(m^2+\nu^2+a^2)^2-4a^2\nu^2},
\nonumber\\ &&M_3=\frac{ma\nu}{m^2+(a+\nu)^2}, \nonumber\\
&&J_1(z_0)=-\frac{\nu z_0}{2}-\frac{ma\nu^2}{m^2+(a-\nu)^2}
+\frac{\nu^2(\nu z_0+ma)}{(z_0+m)^2+(a-\nu)^2}, \nonumber\\
&&J_2=ma
+\frac{2ma\nu^2(m^2+\nu^2+a^2)}{(m^2+\nu^2+a^2)^2-4a^2\nu^2},
\nonumber\\ &&J_3(z_0)=\frac{\nu
z_0}{2}-\frac{ma\nu^2}{m^2+(a+\nu)^2} -\frac{\nu^2(\nu
z_0-ma)}{(z_0+m)^2+(a+\nu)^2}. \label{mj_sup} \eea As before,
$M_1$ and $M_3$ are the total masses of the upper and lower
singular regions, $M_2$ and $J_2$ are the mass and angular
momentum of the central object, while $J_1(z_0)$ and $J_3(z_0)$
denote angular momenta of the parts $0<z\le z_0$ and $-z_0\le z<0$
of the singular regions, respectively. Note that since formulae
(\ref{mj_sup}) have been derived under the supposition
$a^2>m^2+\nu^2$, the limit $a\to 0$ in (\ref{mj_sup}) is
impossible, as $m$ and $\nu$ are real quantities by definition.

It is easy to see from (\ref{mj_sup}) that the aggregate mass of
two singularities always takes a negative value for any positive
$m$ and $\nu\ne 0$:
\be
M_1+M_3=-\frac{4ma^2\nu^2}{(m^2+\nu^2+a^2)^2-4a^2\nu^2}.
\label{m_agg_sup} \ee In figure~5 we have plotted the mass
distributions in the semi--infinite singularities for a particular
choice of the parameters $m=1$, $\nu=2$, $a=3$.

\section{The electrovacuum case}

In the presence of the electric and magnetic charges the general
picture of the singular regions due to the NUT parameter is
similar to that of the vacuum case. However, the singularities
will now be electrically and magnetically charged, and their
masses and angular momenta will have both the gravitational and
electromagnetic contributions. Since the properties of the
semi--infinite singularities are practically independent of the
central body, we shall restrict our consideration to only the
subextreme case defined by the inequality $m^2+\nu^2>a^2+q^2+b^2$.
The symmetry axis then is divided into three regions, as in
figure~1: $\kappa<z<\infty$ (region~I), $|z|<\kappa$ (region~II)
and $-\kappa<z<-\infty$ (region~III). The functions $\omega$,
$\Omega\equiv{\rm Im}\E$, $\Phi$ and $A_3$ which are needed for
the calculation of the Komar quantities show the following
behaviour in the regions I, II, III:

{\it Region I}. For $\rho=0$, $\kappa<z<\infty$ we obtain from
(\ref{DN_Ep}) and (\ref{mf_DN}): \bea &&\omega_{(1)}=2\nu(C_1+1),
\quad \Omega_{(1)}=-\frac{2(\nu z+ma)}{(z+m)^2+(a-\nu)^2},
\nonumber\\ &&\Phi_{(1)}=-\frac{q+\I b}{z+m-\I(a-\nu)}, \quad
A_{3(1)}=-b(C_2+1), \label{axis_em1} \eea where the subindex~1 in
parentheses (and the subindices~2,~3 below) signifies that the
respective function is calculated on the indicated part of the
symmetry axis.

{\it Region II}. When $\rho=0$, $|z|<\kappa$, we have \bea
&&\omega_{(2)}=2C_1\nu+\frac{(\kappa+m)^2+a^2+\nu^2}{a}, \quad
\Omega_{(2)}=-\frac{2\kappa(maz+\kappa^2\nu)}
{(az-\nu)^2+\kappa^2(\kappa+m)^2}, \nonumber\\
&&\Phi_{(2)}=-\frac{\kappa(b-\I q)} {az-\kappa\nu+\I
\kappa(\kappa+m)}, \nonumber\\ &&A_{3(2)}=-C_2
b-\frac{q(\kappa+m)+\nu b}{a} \nonumber\\
&&\hspace{1cm}-\frac{[b(az-\kappa\nu)-\kappa
q(\kappa+m)][(\kappa+m)^2+(a-\nu)^2]}
{a[(az-\kappa\nu)^2+\kappa^2(\kappa+m)^2]}. \label{axis_em2} \eea

{\it Region III}. For $\rho=0$, $\kappa<z<\infty$ one gets \bea
&&\omega_{(3)}=2\nu(C_1-1), \quad \Omega_{(3)}=\frac{2(\nu
z+ma)}{(z-m)^2+(a+\nu)^2}, \nonumber\\ &&\Phi_{(3)}=-\frac{q+\I
b}{z-m-\I(a+\nu)}, \nonumber\\ &&A_{3(3)}=-b(C_2-1)
+\frac{4\nu[q(z-m)-b(a+\nu)]}{(z-m)^2+(a+\nu)^2}. \label{axis_em3}
\eea

The calculation of the masses $M_i$, electric $Q_i$ and magnetic
$B_i$ charges, as well as angular momenta $J_i$ of the
regions~I--III can be carried out with the aid of Tomimatsu's
formulae derived in the paper \cite{Tom3}: \bea M_i&=&M_i^G+M_i^E=
-\frac{1}{4}\int_{d_i}^{u_i}[\omega_{(i)}\Omega_{(i),z}
-2\omega_{(i)}{\rm Im}(\Phi_{(i)}\bar\Phi_{(i),z})]dz \nonumber\\
&&-\frac{1}{2}\int_{d_i}^{u_i} \omega_{(i)}{\rm
Im}(\Phi_{(i)}\bar\Phi_{(i),z})dz \nonumber\\ &=&-\frac{1}{4}
\omega_{(i)}[\Omega_{(i)}(z=u_i)-\Omega_{(i)}(z=d_i)], \nonumber\\
Q_i&=&\frac{1}{2}\omega_{(i)}{\rm Im}
[\Phi_{(i)}(z=u_i)-\Phi_{(i)}(z=d_i)], \nonumber\\
B_i&=&-\frac{1}{2}\omega_{(i)}{\rm Re}
[\Phi_{(i)}(z=u_i)-\Phi_{(i)}(z=d_i)], \nonumber\\
J_i&=&J_i^G+J_i^E= -\frac{1}{8}\int_{d_i}^{u_i}\omega_{(i)}
[2+\omega_{(i)}\Omega_{(i),z} -2\omega_{(i)}{\rm
Im}(\Phi_{(i)}\bar\Phi_{(i),z})]dz \nonumber\\
&&+\frac{1}{2}\int_{d_i}^{u_i} \omega_{(i)}A_3{\rm
Im}(\Phi_{(i),z})dz, \label{Kom_em} \eea where the superscripts G
and E denote the decomposition of masses and angular momenta into
the gravitational and electromagnetic components introduced by
Tomimatsu following Carter's paper \cite{Car}.

The total masses of the regions~I, II, III obtainable with the aid
of the above formulae (\ref{Kom_em}) in which one has to set
$u_1=\infty$, $d_1=u_2=\kappa$, $d_2=u_3=-\kappa$, $d_3=-\infty$,
have the form \bea &&M_1=-\frac{\nu(C_1+1)(\kappa\nu+ma)}
{(\kappa+m)^2+(a-\nu)^2}, \nonumber\\ &&M_2=\left( m+
\frac{2\nu(\kappa\nu+ma)} {(\kappa+m)^2+(a-\nu)^2}\right) \left(
1+\frac{2a\nu(C_1-1)} {(\kappa+m)^2+(a+\nu)^2}\right), \nonumber\\
&&M_3=\frac{\nu(C_1-1)(\kappa\nu-ma)} {(\kappa+m)^2+(a+\nu)^2},
\label{m123_em} \eea and it is easy to verify that the total mass
$M_{tot}$ of the solution is simply
\be
M_{tot}=\sum\limits_{i=1}^3 M_i=m \label{mtotal_em} \ee for all
values of the constant $C_1$. It is worthwhile pointing out that
the cases $C_1=-1$ and $C_1=1$ correspond to vanishing of the
upper and lower semi--infinite singularities, respectively. At the
same time, for instance, the combination of the parameters
$\kappa\nu+ma=0$, $C_1\ne -1$, which also leads to $M_1=0$, does
not annihilate the upper singularity, but only reflects the fact
that the {\it total} gravitational and electromagnetic
contributions into the mass of this singularity are equal in
absolute values and have opposite signs; at the same time, these
contributions calculated for any {\it finite} interval of the
singularity do not give zero.

The distribution of the electric charge is described by the
formulae \bea &&Q_1=-\frac{\nu(C_1+1)[b(\kappa+m)+q(a-\nu)]}
{(\kappa+m)^2+(a-\nu)^2}, \nonumber\\ &&Q_2=\left( q+
\frac{2\nu[b(\kappa+m)+q(a-\nu)]} {(\kappa+m)^2+(a-\nu)^2}\right)
\left( 1+\frac{2a\nu(C_1-1)} {(\kappa+m)^2+(a+\nu)^2}\right),
\nonumber\\ &&Q_3=\frac{\nu(C_1-1)[b(\kappa+m)-q(a+\nu)]}
{(\kappa+m)^2+(a+\nu)^2}, \quad Q_{tot}\equiv\sum\limits_{i=1}^3
Q_i=q, \label{q123_em} \eea whereas for the distribution of the
magnetic charge we obtain \bea
&&B_1=\frac{\nu(C_1+1)[q(\kappa+m)-b(a-\nu)]}
{(\kappa+m)^2+(a-\nu)^2}, \nonumber\\ &&B_2=\left( b-
\frac{2\nu[q(\kappa+m)-b(a-\nu)]} {(\kappa+m)^2+(a-\nu)^2}\right)
\left( 1+\frac{2a\nu(C_1-1)} {(\kappa+m)^2+(a+\nu)^2}\right),
\nonumber\\ &&B_3=-\frac{\nu(C_1-1)[q(\kappa+m)+b(a+\nu)]}
{(\kappa+m)^2+(a+\nu)^2}, \quad B_{tot}\equiv\sum\limits_{i=1}^3
B_i=b. \label{b123_em} \eea One can see that the total charges too
do not depend on the constant $C_1$.

The distribution of the angular momentum in the electrovac DN
solution reminds qualitatively the distribution in the pure vacuum
case, but the corresponding formulae turn out to have a very
cumbersome form and so will be not given here (the reader can work
them out straightforwardly from (\ref{Kom_em})). We only comment
that the choice $C_1=0$ is obligatory for having a finite value of
the total angular momentum, and once this choice is made, the
other constant $C_2$ can be adjusted in such a way that the total
angular momentum be equal to $ma$, in accordance with the
multipole moments (\ref{mult}).

\section{Concluding remarks}

Like in the NUT solution, the semi--infinite singularities in the
combined Kerr--NUT and electrovac DN spacetimes carry infinite
angular momenta, and the only possibility for them to have a
finite total angular momentum is assigning zero value to the
integration constant in the expression for the metric function
$\omega$, in which case the two singularities will possess
oppositely oriented angular momenta cancelling out the infinities
in sum. The total mass of the singularities in this special case
assumes a negative value. In contradistinction to the purely NUT
solution where the counter--rotating singularities and a static
central body form a system which is antisymmetric with respect to
the equatorial plane, in the DN vacuum and electrovac spacetimes
the non--zero parameter $a$ makes impossible such an additional
equatorial symmetry. The analysis carried out in the present paper
and in \cite{MRu} shows how careful one should be when he tries to
establish the multipole structure of spacetimes possessing the NUT
parameter. At the same time, it is surprising how elegantly the
additional parameters $\nu$ and $b$ of the electrovacuum DN
spacetime generalize the known multipole expressions derived for
the Kerr--Newman solution.

\vspace{.5cm}

\noindent{\bf Acknowledgements}

\vspace{.5cm}

\noindent VSM would like to thank the Department of Fundamental
Physics of the Salamanca University where a part of this work was
done for its kind hospitality and financial support. This work was
partially supported by Project 45946--F from CONACyT of Mexico,
and by Project BFM2003--02121 from MCyT of Spain.


\newpage

\begin{figure}[htb]
\centerline{\epsfysize=90mm\epsffile{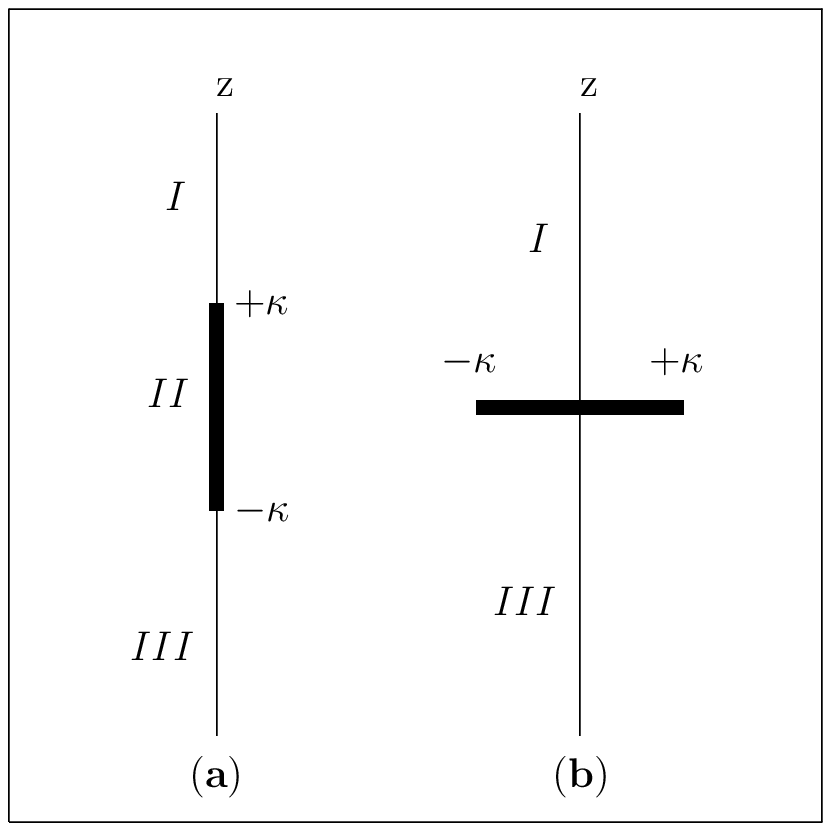}} \caption{Different
parts of the symmetry axis (a) in the case of the subextreme DN
spacetimes; (b) in the case of the superextreme DN spacetimes.}
\end{figure}

\begin{figure}[htb]
\centerline{\epsfysize=80mm\epsffile{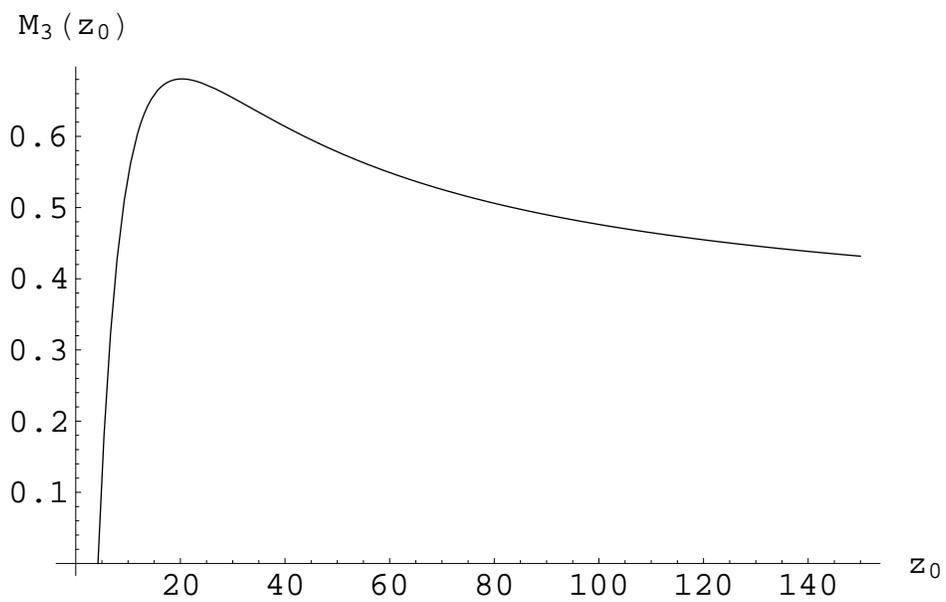}} \caption{The mass
distribution in the semi--infinite singularity of the subextreme
DN solution. The particular choice of the parameters is: $m=5$,
$\nu=3$, $a=4$, $C=-1$.}
\end{figure}

\begin{figure}[htb]
\centerline{\epsfysize=80mm\epsffile{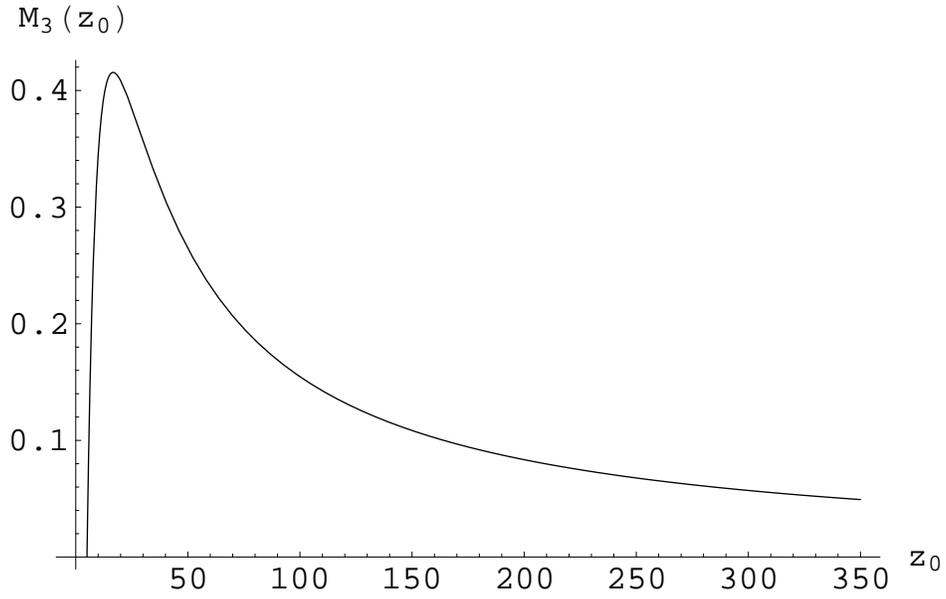}} \caption{The mass
distribution in the semi--infinite singularity of the subextreme
DN solution, the total mass of the singularity being equal to
zero.}
\end{figure}

\begin{figure}[htb]
\centerline{\epsfysize=50mm\epsffile{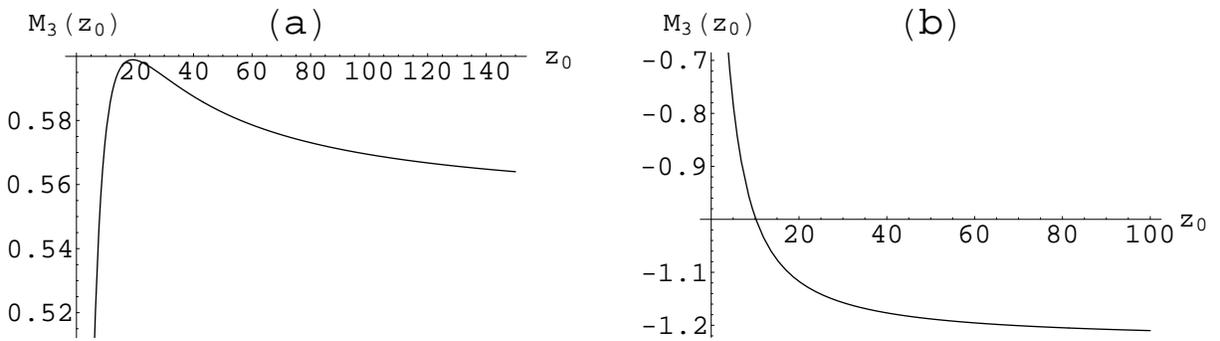}}
\caption{Dependence of the mass distribution in the original DN
semi--infinite singularity on the angular momentum of the central
body.}
\end{figure}

\begin{figure}[htb]
\centerline{\epsfysize=50mm\epsffile{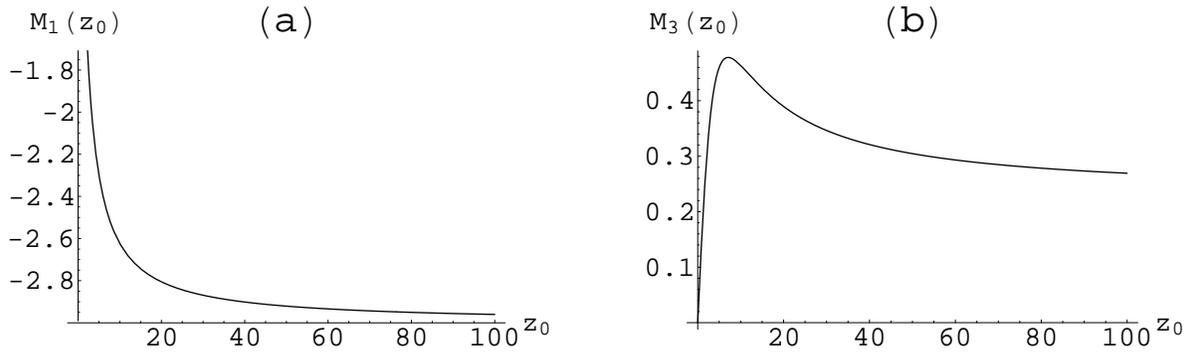}} \caption{The mass
distributions in the upper and lower singularities of the
superextreme DN solution possessing finite total angular momentum
(figures (a) and (b) respectively).}
\end{figure}

\end{document}